\documentstyle{article}
\title{
\begin{flushright}
{\bf\normalsize   COLO-HEP-290}\\
\end{flushright}
\bf Crossover Between Weakly and Strongly Self-avoiding Random Surfaces
}

\author{ {\it C.F. Baillie} \\
         Physics Dept. \\
         University of Colorado\\
         Boulder, CO 80309\\
         USA\\
         \\
         and \\
         \\
         {\it D.A. Johnston}\\
         Dept. of Mathematics\\
         Heriot-Watt University\\
         Riccarton\\
         Edinburgh, EH14 4AS\\
         Scotland}

\begin{document}
  \maketitle
                      {\Large
                      \begin{abstract}
%
We investigate the crossover between weak and strong self-avoidance
in a simulation of random surfaces with extrinsic curvature. We consider
both dynamically triangulated and rigid surfaces with the two possible
discretizations of the extrinsic curvature term.
\\
\\
\\
\\
\\
Submitted to Phys Lett B.
%
                        \end{abstract} }
%
  \thispagestyle{empty}
%
%
  \newpage
%
                  \pagenumbering{arabic}
In recent papers we have investigated the limits of weak \cite{1} and
strong \cite{2} self-avoidance in both fixed and dynamically triangulated
random surfaces. We examined the behaviour of a gaussian plus
extrinsic curvature action of the form
\begin{equation}
Z = \sum_{T} \int \prod_{i=1}^{N-1} d X_i^{\mu} \exp - ( S_g + \lambda
S_{a,e}),
\label{e2}
\end{equation}
where the gaussian action $S_g$ is a straightforward discretization of the
Polyakov string action \cite{3}
\begin{equation}
S_g = {1 \over 2} \sum_{<ij>} (X_i^{\mu} - X_j^{\mu})^2.
\label{e3}
\end{equation}
and the alternative discretizations of the extrinsic curvature
term, usually subscripted ``area'' and ``edge'' respectively can be written as
\begin{equation}
S_a = \sum_i {1 \over \Omega_i } \left( \sum_{j(i) } ( X^{\mu}_i - X^{\mu}_j)
\right)^2
\label{e4}
\end{equation}
for the area term and
\begin{equation}
S_e = \sum_{\Delta_i, \Delta_j} ( 1 - \hat n_{\Delta_i}
 . \hat n_{ \Delta_j} )
\label{e5}
\end{equation}
for the edge term. The sum over triangulations, which in effect makes the
surfaces dynamical,
is omitted in the fixed mesh simulations. At low $\lambda$ surfaces are in a
crumpled phase and at
large $\lambda$ in a smooth phase, with a crumpling transition occurring at
some finite $\lambda$.

An action of the form equ.\ref{e2} was originally suggested both in the context
of QCD strings
and as a potential action for real surfaces such as cell membranes \cite{4}. It
has been
the subject of considerable interest as a candidate discretized string action
\cite{5},
because it appears that $S_g + S_e$ may have a crumpling transition
at which the string tension scales so as to give a non-trivial continuum theory
\footnote{The
action $S_g + S_a$ appears to be afflicted with lattice artifacts \cite{6}.}.
This work has been done {\it without} self-avoidance as it is not relevant from
the string
theory point of view, at least for purely bosonic models. There has been
parallel work in
solid state physics on realistic surface models which are strongly
self-avoiding \cite{7},
to date mostly of fixed (termed ``polymerized'') mesh surfaces. Self-avoidance
can be incorporated into such simulations by surrounding each point on the
discretized mesh with spheres
of diameter $\sigma$
and linking them with tethers that cannot extend beyond some given length $l$.
If $l< \sqrt{3} \sigma$
a sphere, and hence the surface itself, cannot be threaded through three other
spheres and
the surface is said to be strongly self-avoiding - even
distant portions of the surface in the intrinsic metric cannot intersect. If $l
\ge \sqrt{3} \sigma$
intersections are possible, but there is still a local excluded volume effect
due to the spheres, and the surface is said to be weakly self-avoiding.
It is usually more convenient from the computational point of view to divide
the
embedding space into boxes and ensure single occupancy, rather than putting in
the
spheres explicitly.
The current state of
opinion in the solid state physics community, after some initial confusion, is
that
fixed mesh surfaces with strong self-avoidance do not display crumpling
\cite{7}. Our work in \cite{2}
is in agreement with this, finding that the peak in the specific heat that
signals the crumpling
transition is suppressed on fixed meshes by strong self-avoidance. There has
been less work on dynamical
surfaces with strong self-avoidance, but \cite{2} and the independent work of
Kroll and Gompper in \cite{8}
are in broad agreement, finding a {\it branched-polymer} phase at low $\lambda$
rather than a crumpled phase
and a smoother phase at large $\lambda$. It is currently unclear whether there
is a bona fide crumpling
transition here or whether the smooth phase is simply a finite size effect.
In addition \cite{9} claimed to find a crumpled phase for a strongly
self-avoiding dynamical surface at
zero $\lambda$, but their numerical results are, in fact, compatible with those
for a branched polymer.
The only work
on weak self-avoidance has been that in \cite{1}, where we found that weak
self-avoidance had little
effect on the crumpling transition.

We might expect that shortening the lengths of the tethers would allow us to
see the change from this
weak behaviour to the qualitatively different strong behaviour as $l$ passed
through $\sqrt{3} \sigma$.
In this paper we report on the crossover, simulating
sphere diameter $\sigma = 0.25$ and
tether lengths of $0.4,0.433,0.5,0.75,1.0,$
$1.5$ and $2.0$
($l \le 0.433$ is strong) with both the area and edge extrinsic curvature terms
and on fixed and dynamical
surfaces. As our aim is to see the qualitative features rather than obtaining
high numerical accuracy the
bulk of the simulations are carried out on modestly sized surfaces of 72 nodes,
but we
also simulate surfaces of 314 nodes for the case of the edge action, which is
most interesting
from the point of view of continuum string theory, with strong self-avoidance
in order to compare
the results with earlier extensive simulations without self-avoidance \cite{5}.
Again for comparison with earlier work we simulate surfaces of spherical
topology.
We
discuss firstly the edge action on both fixed and dynamical meshes in some
detail, followed
by a more cursory examination of the area action results which are less
interesting from the continuum theory point of view.
For further details on the computational aspects of this work, see
\cite{1,2}.

If the simulations of the solid state physicists are to be believed there
should be no crumpling transition observable
for strongly self-avoiding surfaces with the edge action
on {\it fixed} meshes. As the tether length is relaxed to give weak
self-avoidance
we would expect to see the re-emergence of the crumpling transition between a
low lambda crumpled phase
and a large lambda smooth phase. The crossover between the two types of
behaviour might be expected to occur
in the region of the maximum tether length for strong self-avoidance,
$l=0.433$. These expectations are
confirmed by the results displayed in Fig. 1 for the specific heat, which is
defined as
\begin{equation}
C_{a,e} = { \lambda^2 \over N} ( < S_{a,e}^2> - < S_{a,e}>^2 ),
\label{e6}
\end{equation}
for the area and edge discretizations respectively, and in Fig. 2 for the
gyration radius
\begin{equation}
X2 = { 1 \over 9 N (N -1)} \sum_{ij} \left( X^\mu_i - X^\mu_j \right)^2
q_i q_j.
\label{e7}
\end{equation}
We can see from Fig. 1 that the peak in the specific heat that signals the
crumpling transition is
gradually suppressed as we shorten the tethers. There is no sign of a sharp
change in behaviour
(the results for $l=0.5$ are very similar to those for $l=0.433$, for instance)
though it is
possible that the changeover might be sharper on large meshes. The suppression
is also
evident in Fig. 2 where we see that the increase in the gyration radius that
signals the transition
gradually vanishes as the tether lengths are shortened. We can get
some idea of asymmetries in the configuration of the surface by looking at the
individual components
of the moment of inertia tensor
\begin{equation}
M_{\mu \nu} = {1 \over N^2} \sum_i^N \sum_j^N (X_i^\mu - X_j^{\mu}) (X_i^{\nu}
- X_j^{\nu} )
\label{e12}
\end{equation}
and in this case we find that they remain essentially equal in all the phases,
indicating rough spherical
symmetry. Perhaps the best way to get a qualitative idea of the
surfaces' behaviour is to look at representative ``snapshots'' of the surfaces
rendered in
three dimensional space.
Previous work has shown that without self-avoidance the surfaces look genuinely
crumpled at low lambda
and smooth at large lambda. The behaviour of the weakly self-avoiding surfaces
is similar to this:
in Fig. 3 we render a 72 node weakly self-avoiding surface with $\lambda=0$
which is obviously crumpled,
and in Fig. 4 a weakly self-avoiding surface with $\lambda=3$ (well into the
smooth phase).
As we pass over to strong self-avoidance the typical surface configurations
become similar to
Fig. 5 for both large and small $\lambda$. The surfaces now appear to stay in a
``knobbly''
or crinkled phase, whatever the value of $\lambda$, and the transition is lost.

The behaviour of the edge action with strong self-avoidance on a {\it
dynamical} mesh is radically different. As the self-avoidance
becomes stronger the peak in the specific heat, while suppressed, does not
vanish entirely as can be seen in
Fig. 6. The results for the gyration radius are at first sight rather bizarre,
as it crosses over from
an increase signaling uncrumpling for weak self-avoidance, to
a {\it decrease} at the phase transition for strong self-avoidance, even though
the large $\lambda$ phase
is still smooth. We have plotted this in Fig. 7. Examining
the eigenvalues of the moment of inertia tensor and snapshots of 72 node
surfaces \cite{2} suggested this was the result of
a low $\lambda$ branched polymer phase. A surface
in this phase has two small and one larger eigenvalues of the moment of inertia
tensor, giving a larger
$X2$ than for a crumpled phase, thus explaining the apparently anomalous
results of Fig. 7.  The crossover
between weak and strong behaviour is, as for the rigid surfaces, rather smooth
with no obvious signs of a sudden change at
$l=0.433$.
We can obtain further confirmation of the branched polymer like nature of the
low $\lambda$ phase by looking at
snapshots of the larger 314 node surfaces we simulated. In Fig. 8 we show a
strongly self-avoiding surface with
$\lambda=0.5$ which clearly displays the branching tendency of the surface.
This persists up to the transition point, signaled by the peak in the
specific heat, above which the surfaces appear largely smooth.

The area discretization would appear to be of less interest as a potential
candidate for a continuum theory
because  of the problems with lattice artifacts alluded to earlier \cite{6}. We
have, however, simulated this
too for completeness, finding that the specific heat, which displays a bump
rather than a sharp peak,
is largely unchanged by increasing the strength of the self-avoidance and that
the increase in $X2$ with
$\lambda$ is suppressed as the self-avoidance becomes stronger on both fixed
and dynamical meshes.

We can summarize our results for the edge discretization as follows:
\begin{itemize}
\item{} The crossover from weakly to strongly self-avoiding behaviour appears
to be rather smooth.
\item{} On strongly self-avoiding fixed meshes snapshots suggest that only a
uniform ``knobbly'' phase remains.
\item{} On strongly self-avoiding dynamical meshes there is a low $\lambda$
branched polymer phase and a large $\lambda$ smooth phase
\end{itemize}
It would thus appear that strong self-avoidance has a radically different
effect on fixed meshes, which might
be considered models of polymerized surfaces, and dynamical meshes, which might
be considered models
of fluid surfaces.
It would be interesting to conduct simulations of larger surfaces at higher
statistics to see if the qualitative
picture for the various phases on both fixed and dynamical meshes suggested by
the current work is
supported by firmer numerical evidence and to compare more closely with the
solid state physics
simulations which have largely been with surfaces of planar topology.

This work was supported in part by NATO collaborative research grant CRG910091.
CFB is supported by DOE under contract DE-AC02-86ER40253 and by AFOSR Grant
AFOSR-89-0422.
The computations were performed on the TC2000 Butterfly and
the Sequent Symmetry at Argonne National Laboratory, the GP1000
Butterfly at Michigan State University, and the Myrias SPS-2
at the University of Colorado. We would like to thank
R.D. Williams for help in developing initial versions of the dynamical
mesh code.

\vfill
\eject

\vfill
\centerline{\bf Figure Captions}
\begin{description}
\item[Fig. 1.]
The specific heat $C_e$ versus $\lambda$ on a fixed mesh.
\item[Fig. 2.]
The gyration radius $X2$ versus $\lambda$ for the edge curvature on a fixed
mesh.
\item[Fig. 3.]
Weakly ($l=2$) self-avoiding 72-node fixed surface with $\lambda=0$.
\item[Fig. 4.]
Weakly ($l=2$) self-avoiding 72-node fixed surface with $\lambda=3$.
\item[Fig. 5.]
Strongly ($l=0.433$) self-avoiding 72-node fixed surface with $\lambda=3$.
\item[Fig. 6.]
The specific heat $C_e$ versus $\lambda$ on a dynamical mesh.
\item[Fig. 7.]
The gyration radius $X2$ versus $\lambda$ for the edge curvature on a dynamical
mesh.
\item[Fig. 8.]
Strongly ($l=0.433$) self-avoiding 314-node dynamical surface with
$\lambda=0.5$.

\end{description}
\end{document}